\documentclass[fleqn,usenatbib]{mnras}

\usepackage{newtxtext,newtxmath}

\usepackage[T1]{fontenc}

\DeclareRobustCommand{\VAN}[3]{#2}
\let\VANthebibliography\thebibliography
\def\thebibliography{\DeclareRobustCommand{\VAN}[3]{##3}\VANthebibliography}

\usepackage{graphicx}	
\usepackage{amsmath}	
\usepackage{tabularx}
\usepackage{lscape}
\usepackage{rotating}

\newcommand{\maxi}{\emph{MAXI} }
\newcommand{\nicer}{\emph{NICER} }

\title[]{Spectral and timing evolution of MAXI J1631--479 during the 2018-19 outburst with \nicer}

\author[Rout et al.]{
Sandeep K. Rout,$^{1,2}$\thanks{E-mail: skrout@prl.res.in (SKR)}
Mariano M\'endez,$^{3}$
Tomaso M. Belloni,$^{4}$
Santosh Vadawale$^{1}$
\\
$^{1}$Physical Research Laboratory, Navarangpura, Ahmedabad 380009, Gujarat, India\\
$^{2}$Indian Institute of Technology, Gandhinagar, Gujarat, India\\
$^{3}$Kapteyn Astronomical Institute, University of Groningen, PO Box 800, NL-9700 AV Groningen, the Netherlands\\
$^{4}$INAF-Osservatorio Astronomico di Brera, via E. Bianchi 46 I-23807, Merate, Italy
}

\date{}

\pubyear{2021}

\begin{document}
\label{firstpage}
\pagerange{\pageref{firstpage}--\pageref{lastpage}}
\maketitle

\begin{abstract} 
The X-ray transient MAXI J1631--479 went into outburst on 2018 December 21 and remained active for about seven months. Owing to various constraints it was monitored by \nicer only during the decay phase of the outburst for about four months. The \nicer observations were primarily in the soft state with a brief excursion to the hard intermediate state. While the soft state spectrum was dominated by thermal disk emission, the hard intermediate state spectrum had maximum contribution from the thermal Comptonization. Almost all intermediate-state power spectra had a Type-C low frequency quasi-periodic oscillation (within 4 - 10 Hz), often accompanied by a harmonic component. The frequency of these oscillations increased and the fractional rms decreased with inner-disk temperature suggesting a geometric origin. One observation in the middle of the outburst during the hard intermediate state had two non-harmonically related quasi-periodic oscillations. The rms versus QPO frequency relation suggests that while the higher frequency peak is definitely of Type-C, the lower frequency peak could be Type-B in nature. The rms spectra during the intermediate state had a hard shape from above 1 keV. Below 1 keV the shape could not be constrained in most cases, while only a few observations showed a rise in amplitude.

\end{abstract}




\section{Introduction}

The outburst of a black-hole binary (BHB) is characterised by various states depending on the spectral and temporal properties. Since the discovery of X-ray transients more than four decades ago, the state classification schemes have undergone much sophistication matching with the development in instrumentation \citep{miyamoto93, remillard06, done07, belloni11}. With an intent to incorporate all the observable properties, a typical BHB outburst can be categorised into four states - the low hard state (LHS), the high soft state (HSS), and the hard- and soft- intermediate states (HIMS and SIMS) \citep{homan05,belloni10,belloni16}. These states can be identified by locating the source in the hysteresis tracks of the hardness-intensity diagram \citep[HID;][]{homan01,belloni10} and the total rms - intensity diagram \citep[RID;][]{munozdarias11}. The hardness - rms diagram \citep[HRD;][]{belloni10} also serves as a useful tool in this regard. During an outburst, a BHB rises from quiescence in the LHS when the accretion disk is truncated at a large distance and thermal viscous instability kicks in \citep{lasota01}. In the LHS, the energy spectrum is hard and is dominated by power law emission, due to inverse Comptonization, with negligible contribution from the thermal disk \citep{gilfanov10}. The power density spectrum (PDS) displays a flat-top band limited noise and is marked by maximum variability ($30-40$ per cent). Strong Type-C quasi periodic oscillations (QPO), often accompanied by harmonics, are also a characteristic of the PDS during LHS \citep{casella04,casella05,belloni10}. From here the hardness decreases as the contribution of the thermal disk starts to dominate the spectrum. This is believed to happen as the inner radius of the accretion disk moves inward, driven by an increase in accretion rate \citep{done07}. This track is marked by a passage of the source to the HSS through the HIMS and SIMS. The intermediate state have contributions from both the thermal disk and power law with the HIMS being slightly harder than the SIMS \citep{gierlinski05}. Thus, it is difficult to distinguish the two from the HID and energy spectrum. A clear distinction between the two intermediate states can be drawn from the variability properties. While the HIMS is an extension of the LHS with fractional rms around $\sim 10$ per cent, the SIMS has much lesser rms (few percent) and is often marked by Type-B QPOs \citep{casella05,belloni16}. The HSS is the softest state with maximum contribution from thermal disk emission and minimum variability ($\sim 1$ per cent or less), with occasional QPO detection \citep{gilfanov10}. With decrease in luminosity the source moves from HSS back to the LHS via the same intermediate states. Occasionally, the source also undertakes excursions to the intermediate and some anomalous states \citep{belloni16}.     

\begin{figure*}
\centering
\includegraphics[scale=.5]{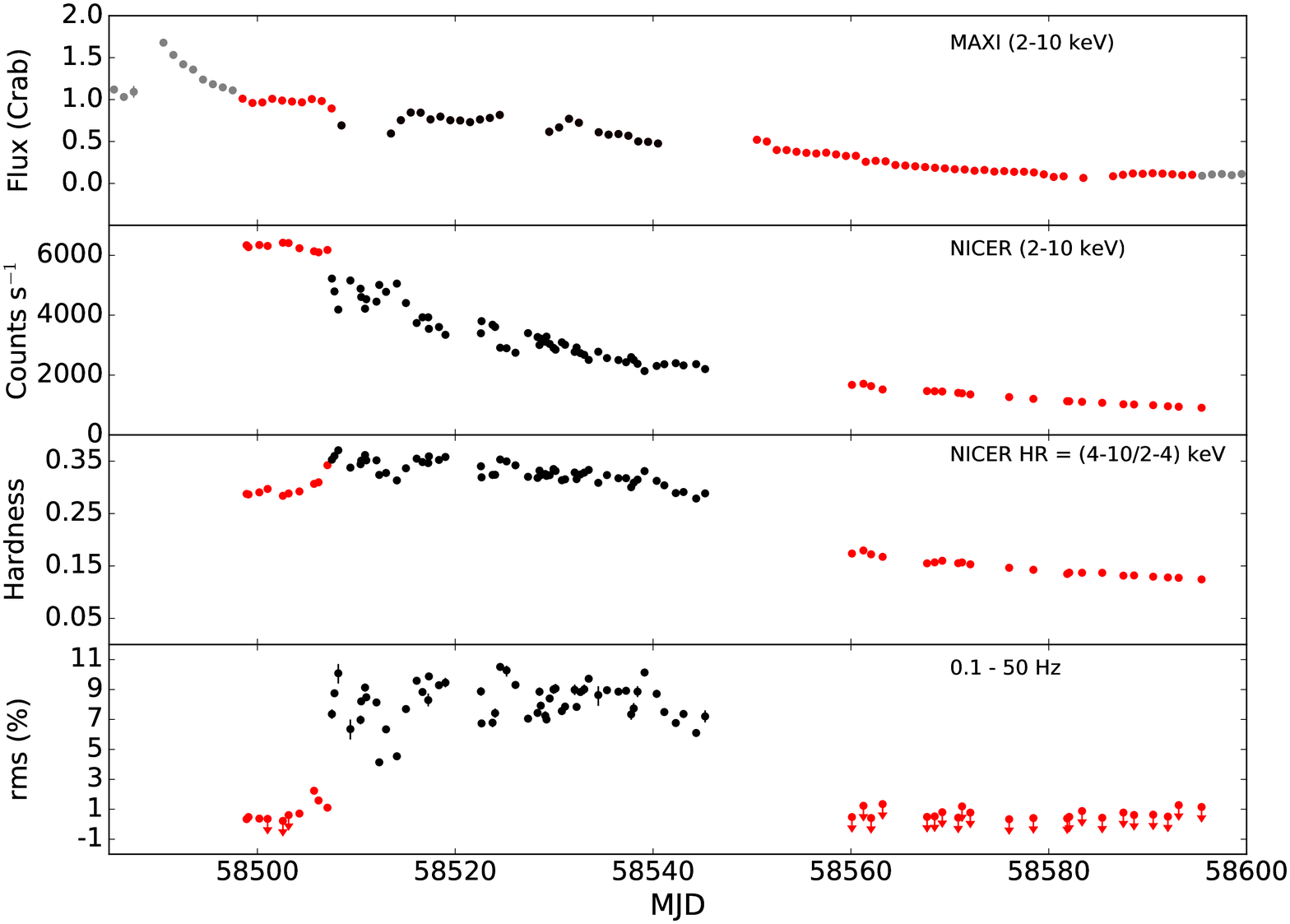} \label{timana}
\caption{Top row: \maxi light curve in the 2 - 10 keV range. Secod row: \nicer light curve in the same energy range. Third row: Time evolution of hardness with \nicer. Fourth row: Time evolution of broadband fractional rms measured with \nicer. The red and black colors, in this and the other figures, represent HSS and HIMS respectively. The grey dots in the \maxi light curve are the points not observed by \nicer.}
\label{fig:timevo}
\end{figure*}

While the phenomenology of QPOs is fairly well understood, their origin is still debated. various models that attempt to explain the phenomenon of QPOs can be broadly divided into two categories - 1) based on intrinsic variability in the plasma and 2) due to geometric effects. \citet{tagger99} proposed the Accretion-Ejection Instability model in which spiral wave instabilities in the density and scale height of a thin accretion disk results in standing wave patterns that form LFQPOs. According to \citet{molteni96} and \citet{chakrabarti08} LFQPOs can be formed due to oscillations of the shock location in the two-component accretion flow model. Pressure wave oscillations within the boundary of Corona can also lead to resonant modes that modulate the Compton upscattered photons resulting in LFQPOs and the associated noise component \citep{cabanac10}. Apart from these, several other attempts have been made to explain the QPO phenomena due to some sort of oscillation or variability in the accretion flow \cite[eg.][]{wagoner01, titarchuk04, oneill11}. Time-dependent Comptonization models were developed to explain the various phenomenology of kilo-Hertz QPOs in neutron star low mass X-ray binaries \citep{lee98,kumar14,karpouzas20}. Recently, \citet{garciaf21} extended this model to explain the rms amplitude and phase lag spectra of Type-B QPOs in the BHB MAXI J1348–-630 by incorporating two separate Comptonization regions. Under the assumption that there is a dynamical mechanism that excites the oscillations, \citet{garciaf21} were able to explain the radiative properties, rms and lag spectra, of the Type-B QPO in this source. While most of the models based on intrinsic variability are quite different from each other, the other class of models based on geometric effects attempt to explain the QPOs with the phenomenon of Lense-Thirring precession. \citet{stella98} and \citet{stella99} proposed the relativistic precession model (RPM) in which QPOs are formed due to Lense-Thirring precession at some characteristic radius which decides the frequency. \citet{ingram09} extended the model under a truncated disk framework to explain the QPOs and other noise elements by precession of a hot inner flow (Corona). \citet{schnittman06} also proposed a slight variation of the RPM where a precessing ring of matter leads to modulation of X-rays. Although, ascertaining a particular model for the QPOs is still debated, several studies done recently have strongly preferred a geometric origin over intrinsic variability. \citet{motta15, heil15, vandeneijnden17} have shown that the Type-C QPOs have a significant inclination dependence in their strength. They also suggest that Type-B QPOs likely have a different origin than Type-C. \citet{ingram16, ingram17} went on to carry out phase-resolved spectroscopy of QPOs to verify that the reflection spectrum varies with different phases of the QPO providing strong evidence for geometric origin. 

MAXI J1631--479 (hereafter J1631) is a newly discovered galactic X-ray transient which hosts a rapidly spinning massive black hole at its center \citep[][Rout \& Vadawale 2021 (under review)]{xu20}. During the HSS, J1631 exhibits outflow of matter in the form of ultra-fast disk winds. It also shows radio flaring during state transitions and a radio/X-ray luminosity pattern typical of BHBs \citep{monageng21}. \citet{fiocchi20} and Rout \& Vadawale (2021) have reported on the presence of hybrid plasma in the \emph{INTEGRAL} and \emph{NuSTAR} spectra respectively.  
In this work we carry out a comprehensive spectral and timing analysis of the X-ray transient J1631 using data from \nicer. We endeavour to understand the phenomenology of state transition and the associated time variability, especially the rms spectra and low-frequency QPOs.

\begin{figure*}
\centering
\includegraphics[scale=0.5]{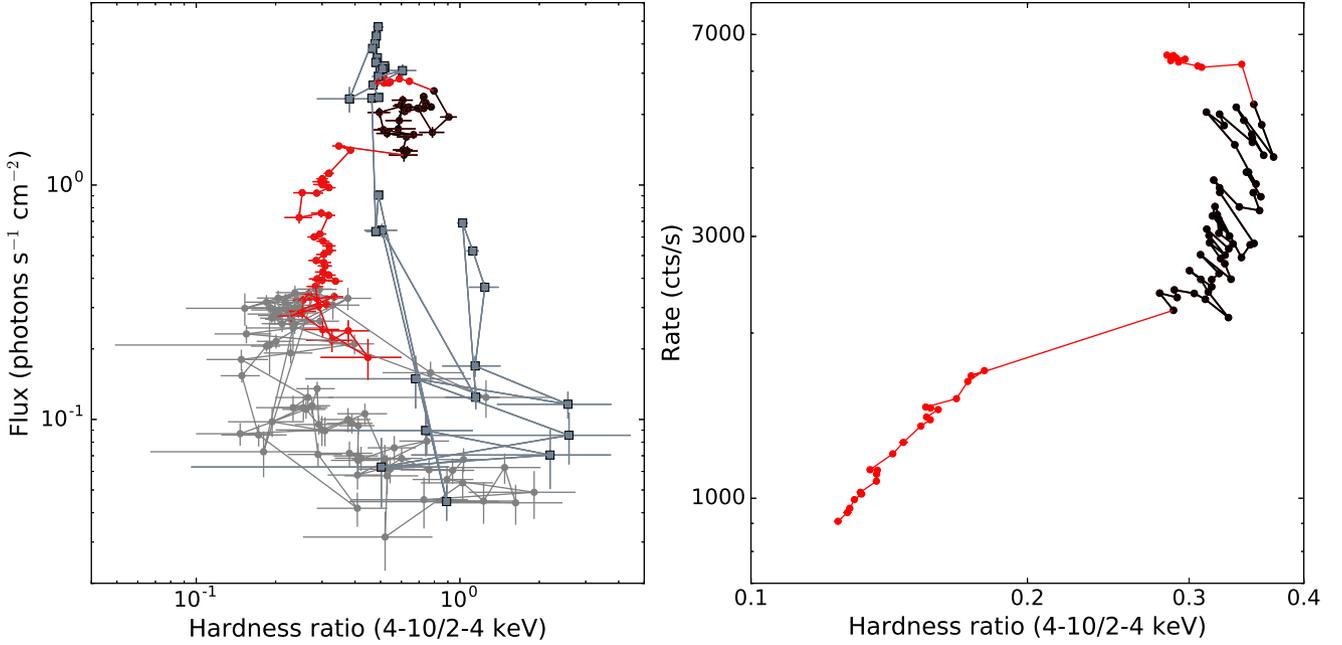}
\caption{Hardness-intensity diagrams with \maxi (left panel) and \nicer right panel). The grey points in the \maxi HID are those that do not have a corresponding point in \nicer. The initial points in the \maxi HID are marked with dark grey squares to distinguish from the final points.}
\label{fig:hid}
\end{figure*}

\section{Observation and Analysis}

\subsection{Data Reduction}

J1631 went into outburst on 21 December 2018 and was first detected with \textit{MAXI}/GSC. After an ambiguity due to its proximity with the pulsar AX J1631.9--475, located within the \textit{MAXI} error circle, it was later confirmed to be a new X-ray transient using $\sim 14$ ks data from \textit{NuSTAR} on 28 December \citep{miyasaka18}. Due to Sun-angle constraints, the peak of the outburst was missed by \textit{NICER} as well as most other X-ray satellites. From 2019 January 15 onward J1631 was observed with \textit{NICER}, almost daily, with exposures ranging from a few hundred seconds to a few thousand seconds till 2019 July 4. All the analysis in this work were done using \texttt{HEASoft-v6.25}. Analysis of the X-ray Timing Instrument \citep[XTI;][]{gendreau12} data was carried out using \texttt{NICERDAS-v5.0}. After generating the event files using \texttt{nicerl2}, the higher level products were generated with the ftool \texttt{xselect}. Several observations in the beginning were affected by flaring events. These were removed by setting \texttt{COR\_SAX} $>\, 6$. The intervals with count rate greater than 1 counts s$^{-1}$ in 13 - 15 keV range were also rejected as they are affected by background flares. The background spectra for each observation were generated using \texttt{nibackgen3C50-v5}. Each \emph{NICER} observation has several data segments based on the ISS orbit. If the count rate in the individual segments differed significantly (i.e., $\gtrapprox 400$ counts s$^{-1}$), they were analysed separately. After 2019 March 20, the background starts to dominate beyond 8 keV and spectral analysis was carried out in the range of 0.5-8 keV. Further, from April 22 the background was dominant even below 8 keV. Thus, in this paper we only present the results till 2019 April 21. This amounts to a total of 88 observations spanned between 2019 January 15 (OBSID: 1200500101) to 2019 April 21 (OBSID: 2200500131). 

\subsection{Timing Analysis} \label{timanal}

Time-series analysis was done using the \texttt{GHATS} package \footnote{http://www.brera.inaf.it/utenti/belloni/GHATS\_Package/Home.html}. \textit{NICER} data were rebinned by a factor of $10^4$, which brought down its time resolution to 400 $\mu$s, corresponding to a Nyquist frequency of 1250 Hz. PDS in the energy range of $0.5-10$ keV were constructed for a time series consisting of 65536 time bins. Each series was, thus, $\sim 26.21$ s long and led to a minimum frequency of $\sim 0.04$ Hz. All PDS in a segment were then averaged and rebinned logarithmically. We have verified that the source shows very little variability for frequencies below 0.04 Hz and the choice of a longer time series would result in decrease in their number which would lead to poor statistics for the averaged PDS. The average power in the 100 - 1250 Hz, where the source showed no intrinsic variability,  was subtracted from the Leahy normalized PDS which were subsequently converted to squared fractional rms \citep{belloni90}. The power spectral analysis was done with raw counts without subtracting the background. The PDS during the HIMS were then fitted by a multi-Lorentzian model \citep{belloni02}. Depending on the shape and number of QPOs, the PDS required about 2 to 5 Lorentzians. The multi-Lorentzian model provided a satisfactory fit for all the PDS. During the HSS, the PDS required only a single power law to be fitted reasonably.

To construct rms spectra, the above method was repeated for the following energy bands: E1:$0.2-0.8$, E2:$0.8-1.0$, E3:$1.0-2.0$, E4:$2.0-6.0$, and E5:$6.0-12.0$ keV. These intervals were selected so as to ensure that each of the energy bands receive roughly similar throughput. The fractional rms in the $0.1-50$ Hz range was calculated by integrating the total variability from the best-fit multi-Lorentzian model. The rms was then plotted as a function of the above defined energy bands. 

\subsection{Spectral Analysis} \label{specana}

Spectral analysis was carried out using \texttt{xspec-v12.10.1} \citep{arnaud96}. The background spectrum for each observation was generated using the tool \texttt{nibackgen3C50}. In order to avoid over-sampling the spectra were rebinned by a factor of 3 and ensured that each spectral bin had a minimum of 30 counts. A systematic error of 1 per cent was added to all the channels. The two main components of a black-hole binary spectrum are a multi-color black body \citep{matsuoka09} and a thermal Comptonization component \citep{zdziarski96,zycki99}. Therefore, we fitted the $0.5-10$ keV \emph{NICER} spectra with the \texttt{xspec} model - \texttt{TBabs*(diskbb+nthcomp)}. The solar abundances in \texttt{TBabs} was set according to \citet{wilms00} and the cross-sections were taken from \citet{verner96}. Residuals around $\sim 2$ keV revealed complex features which are caused due to calibration uncertainty at the Si and Au edges. We added two \texttt{gaussian}s at $\sim 1.7$ and $\sim 2.2$ keV. Another bump at $\sim 1.2$ keV was also apparent, which was also due to calibration uncertainty at the low energy tails and became visible owing to the high absorption suffered by the source. The best fit hydrogen column density ($N_H$) was found to be $\sim 6 \times 10^{22}\, cm^{-2}$, albeit, with some variability primarily because of degeneracy with \texttt{diskbb} norm. The average of all best fit $N_H$ was found to be $6.34 \times 10^{22}\, cm^{-2}$.  In most observations during the bright phase of the outburst, an emission feature was apparent at the Fe K$\alpha$ range of $6-7$ keV. So, another \texttt{gaussian} was added to each spectrum. The strength of this line varied across the observations. In some occasions the line energy pegged at $6.4$ keV suggesting that the \texttt{gaussian} is, perhaps, not the best model for a relativistically skewed line. However, a more sophisticated reflection spectroscopy was not the plan of this paper and hence we did not pursue it. During the HSS, on 2019 January 16, the residuals indicated a dip at $\sim 7.2$ keV. A negative \texttt{gaussian} was added to the model to account for the dip. The best fitting line energy was $\sim 7.4$ keV with an FWHM of $\sim 0.1$ keV. These features are an indication of absorption by blue-shifted H- or He-like Fe ions originating from an equatorial disk wind and have also been detected by \citet{xu20}.  

Although a reasonably good fit was obtained using the above procedure, the instrumental \texttt{gaussian}s seemed to over fit the continuum. There was also a clear zig-zag pattern in the residuals between $1-2.5$ keV. Therefore, we re-fitted all the spectra in the range of $2.5-10$ keV avoiding the instrumental features. The best-fit parameters and fluxes were very similar to the previous case, the average $N_H$ being $6.44 \times 10^{22}$ cm$^{-2}$. This suggests that both ranges could be used for studying the spectral properties but, we chose the later range of $2.5-10$ keV in this work. To avoid the effect of the correlation between $N_H$ and \texttt{diskbb} norm on the parameters, $N_H$ was fixed to the average value of $6.44\times 10^{22}$ for all observations.    

\begin{figure}
\centering
\includegraphics[width=.95\columnwidth]{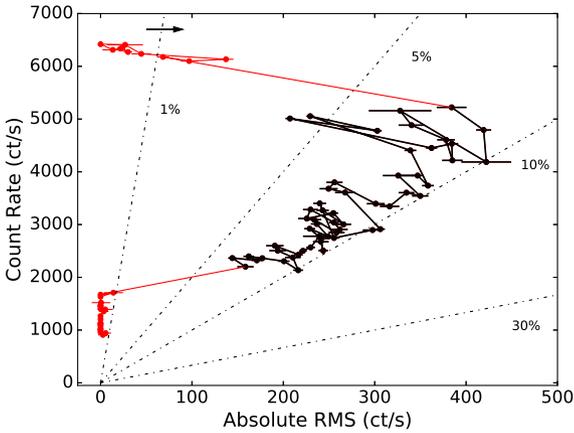}
\caption{RMS - Intensity diagram (RID). The source begins from low variability soft states and transitions to intermediate variability in the HIMS. Finally, it moves back to the low variability soft state till the last observation. }
\label{fig:rid}
\end{figure}

\section{Results}

\subsection{Outburst evolution}

The \emph{MAXI} light curve of J1631 is shown in the top panel of the Figure \ref{fig:timevo}. The peak of the outburst is on 2019 January 07, following which the flux continuously decayed. As mentioned before, \emph{NICER} started observing from 2019 January 15. \emph{NICER} light curve and hardness ratio curve are displayed in the second and third panels of Figure \ref{fig:timevo}. \emph{NICER} started observing the source when it was in the HSS (shown with red colour). On 2019 January 24 J1631 transitioned into a harder state which was marked by a drop in count rate by about $\sim 1000\, counts/s$ and an increase in hardness. The transition is even more clearly seen in the increase of the broadband rms from about 2 per cent to 7 per cent as shown in the bottom panel of Figure \ref{fig:timevo}. From a timing and spectral study, we find that the source transitioned into the HIMS and remained there till 2019 March 03 (shown in black circles in the figure). Following this date, a long data gap ensued and when the source was re-observed it had transitioned back to the HSS. 

The left panel of Figure \ref{fig:hid} shows the HID using \maxi light curve. The outburst began in a hard state with the hardness ratio being $\sim 1$ till 2018 December 31. By 2019 January 07, the source had reached the peak and was in a soft state (hardness ratio $\sim 0.5$). As the decay began, J1631 remained in the HSS till January 24 when it transitioned to the HIMS (shown with black points). The excursion to HIMS lasted for a little more than a month following which the source transitioned back to HSS. During the decaying HSS, the hardness was stable around 0.2 while the flux decreased by an order of magnitude. Finally, the source transitioned to the LHS at very low count rates. The \nicer HID (plotted in the right panel of Figure \ref{fig:hid}) is more clear, but is sparse given that the observations did not cover the entire outburst. \nicer began the observations on January 15 when J1631 was in the HSS and continued through the transition to the HIMS. The HIMS observations were abruptly halted on March 3 and when the observations resumed the source was already in the HSS. A peculiar feature of the \nicer HID is that during the prolonged excursion to the HIMS the count rate decreased by a large factor from $\sim 5500$ counts/s to $\sim 2000$ counts/s giving it a vertical shape. Generally, these excursions occur for shorter periods and at a similar flux level.   

\begin{figure*}
\centering
\includegraphics[scale=.4]{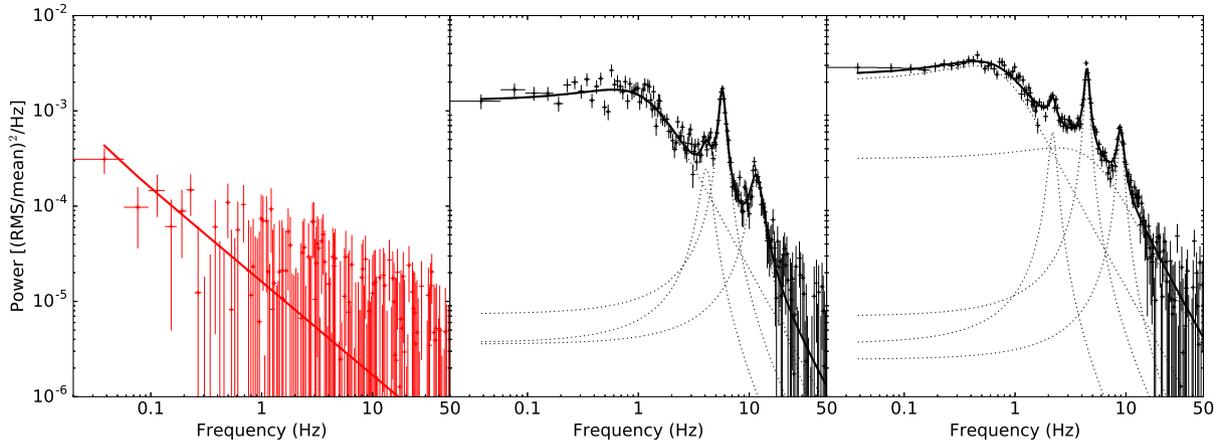}
\caption{From left to right: \nicer PDS observed on 2019 January 17, March 26, and April 16.}
\label{fig:pds}
\end{figure*}

\begin{figure}
\centering
\includegraphics[width=.95\columnwidth]{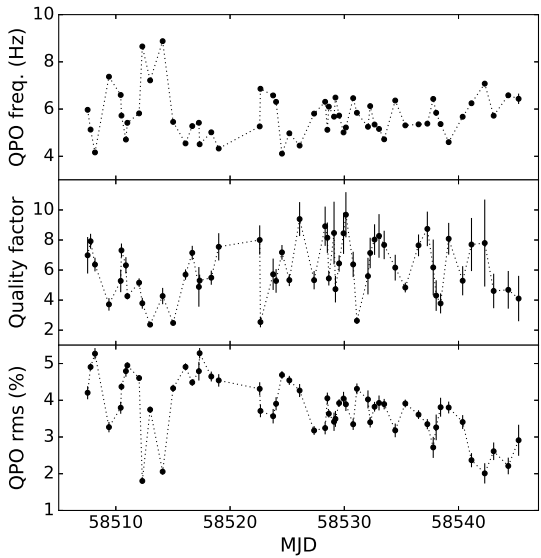}
\caption{From top to bottom: Variation of QPO frequency, quality factor and QPO rms for all observations in the HIMS}
\label{fig:qpoevo}
\end{figure}

\begin{figure*}
\centering
\includegraphics[scale=.5]{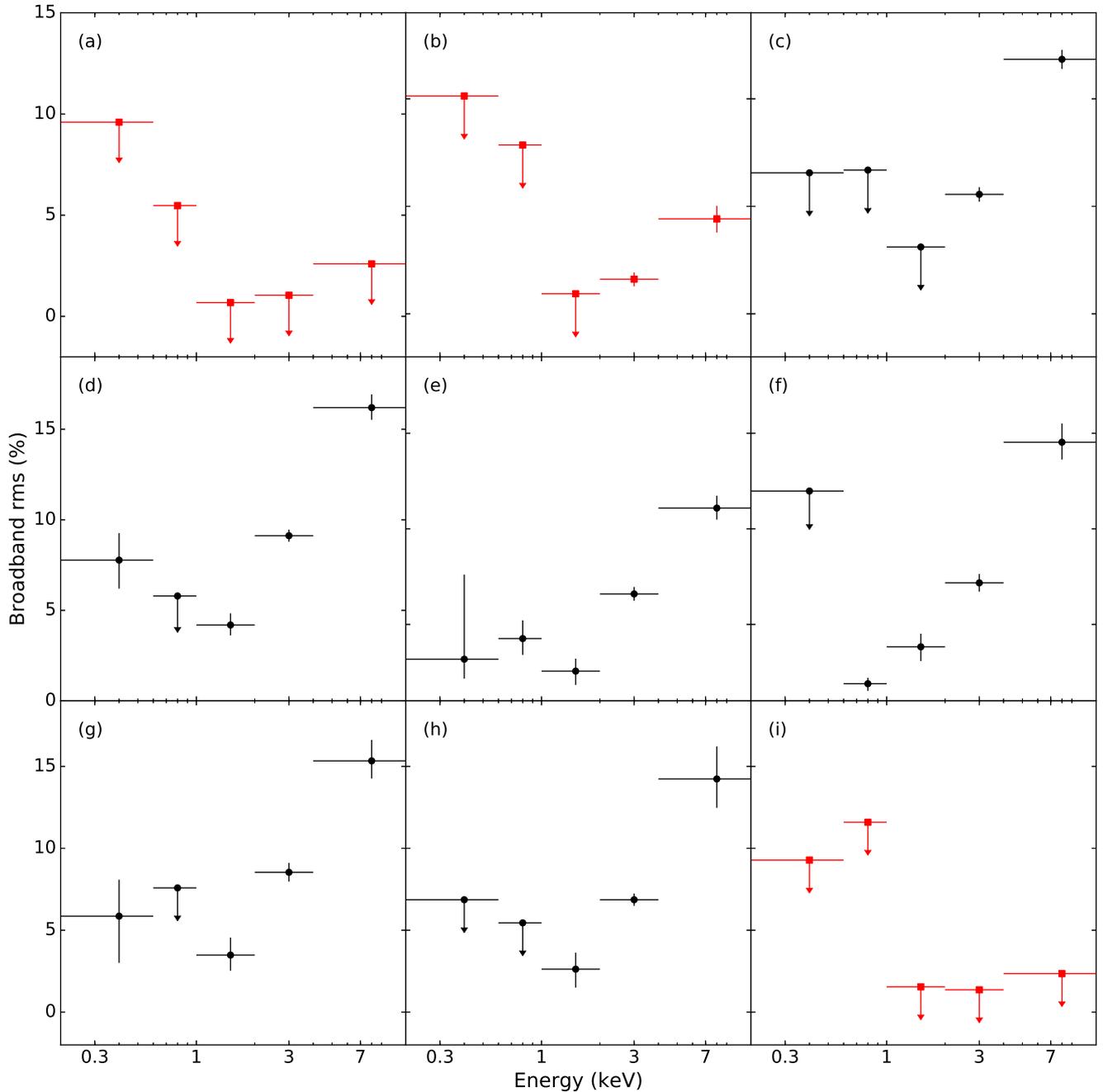}
\caption{Characteristic rms spectra spaced across the entire outburst. From panels (a) to (i): 2019 January 20, January 22, January 26, February 2, February 13, February 17, February 21, March 1, March 20}
\label{fig:rmsspec}
\end{figure*}

\subsection{Timing properties}

The state classification for J1631 is confirmed by studying the time variability (see Figure \ref{fig:rid}). The HSS was marked by very low variability ($\sim 1$ per cent broadband rms), while rms in the HIMS was in the range of $8-10$ per cent. Apart from the broadband rms, all the PDS during the HIMS were characterised by a flat-top noise and a Type-C QPO whose centroid frequency roughly coincided with the break frequency of the red noise. Figure \ref{fig:pds} shows three characteristic PDS spaced across the outburst. The PDS during the HSS (2019 January 17) shows $0.38 \pm 0.05$ per cent variability and contains significant power upto only $\sim 0.3$ Hz. All the PDS during the HSS have a similar power distribution and were fitted by a power law model. The second and third panels display PDS in the HIMS measured on 2019 January 27 and February 11 respectively. The right-hand panel shows the only PDS which has a sub-harmonic QPO. The sub-harmonic and second harmonic QPOs have a frequency of $2.19 \pm 0.03$ and $8.95 \pm 0.04$ Hz which are well-placed compared to the fundamental QPO at $4.45\pm 0.01$ Hz. Most of the other PDS only have the second harmonic and some PDS do not even have any harmonic component. 

The middle panel shows the PDS which displays a QPO-like feature below the fundamental frequency at a non-harmonic ratio. While the fundamental and the second harmonic are at $5.72 \pm 0.02$ Hz and $11.54 \pm 0.10$ Hz respectively, this non-harmonic feature is detected at a frequency of $4.09 \pm 0.08$ Hz and has a quality factor of 4.06. Although these properties qualify it to be considered as a QPO, we remark that the addition of that Lorentzian component accounts for the extra power around 4 Hz, but no clear peak is visible in the PDS, making the identification as a 4 Hz QPO uncertain. 

The time evolution of the fundamental QPO and its properties, during the HIMS, is shown in Figure \ref{fig:qpoevo}. The QPO frequency laid, mostly, within $4-7$ Hz going beyond 8 Hz on two occasions. The quality factor varied between 2 and 10 throughout the observations. The QPO rms showed a declining trend, starting from 4 - 5 per cent during the beginning of HIMS and reaching 2 - 3 per cent towards the end.        

Nine representative rms spectra in the broadband frequency range of $0.1-50$ Hz are displayed in Figure \ref{fig:rmsspec}. The \nicer energy range was divided into 5 energy bands (as spelled in Section \ref{timanal}) and the rms was calculated by fitting the PDS in each of these bands. The rms of almost none of the energy bands could be constrained during the HSS. It was only during 2019 January 22 (Panel b) that the rms of the last two bands were constrained. Since all other spectra in the HSS only have upper limits, it would be difficult to infer much from them. The rms spectra in the HIMS were characterised by a hard shape with the rms rising from to E3 to E5. Here also the first two bands could not be constrained for most of the observations, possibly owing to low statistics. However, in a few occasions, as shown in Panels (d), (e), (f), and (g), one or both of the E1 and E2 bands were constrained. If these were to represent the entire ensemble, it could be inferred that the rms in the first two bands (E1 and E2) would be slightly higher than E3 with the entire spectrum having a concave shape.

\begin{figure}
\centering
\includegraphics[width=.95\columnwidth]{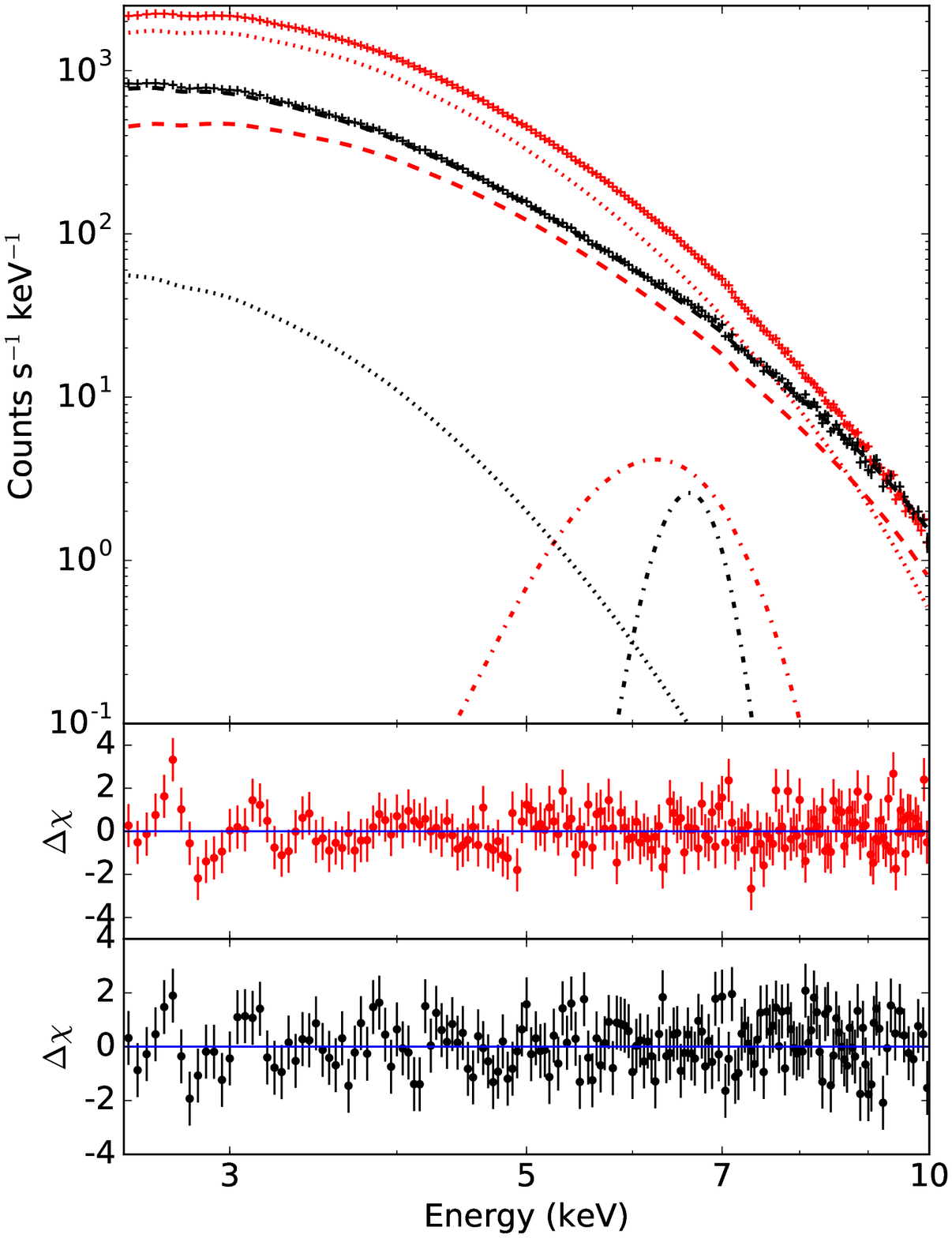}
\caption{Top panel: Two characteristic spectra representing HSS (red, 2019 January 18) and HIMS (black, 2019 February 28), along with individual model components. The dotted line represents the thermal disk component and the dashed line represents the thermal Comptonization component. The gaussians for Fe line are plotted with dot-dashed lines. The bottom two panels show the best fit residuals.}
\label{fig:specres}
\end{figure}

\subsection{Spectral properties}

\begin{figure*}
\centering
\includegraphics[scale=.5]{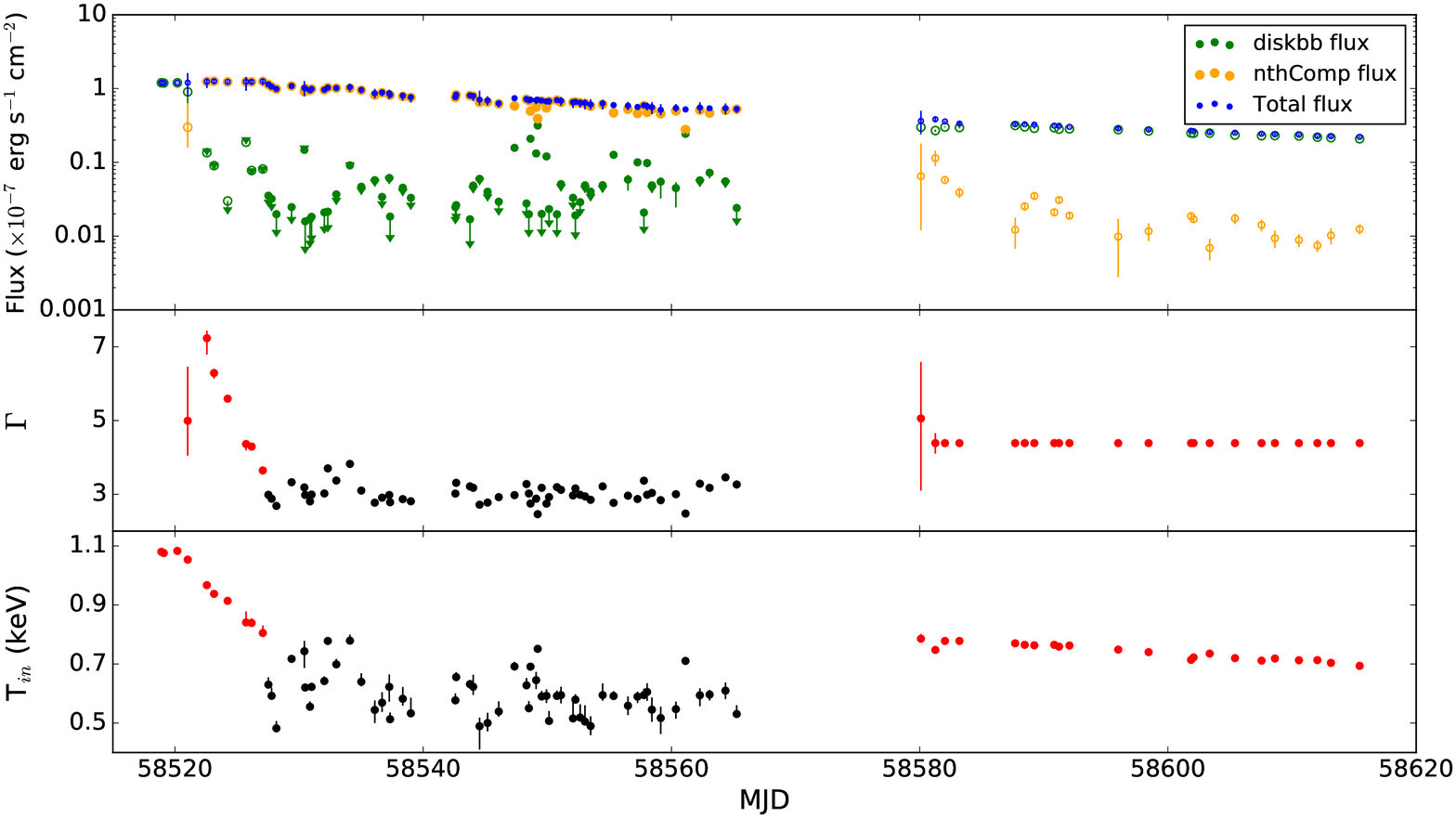}
\caption{Top panel: Time evolution of unabsorbed flux along with the two primary contributors. Middle panel: Evolution of photon index. After January 20, the index was fixed to 4.39 as it could not be constrained. Bottom panel: Variation of inner-disk temperature. In the last two panels, the red and black colors represent HSS and HIMS respectively.}
\label{fig:specevo}
\end{figure*}

As mentioned in Section \ref{specana}, the spectral fitting was done with the model \texttt{TBabs*(diskbb + nthComp + gaussian)}. The seed photon temperature in \texttt{nthComp} was tied to the inner-disk temperature in \texttt{diskbb}. The \texttt{inp\_type} parameter was fixed to 1, thus, assuming a disk blackbody for seed photons. None of the observations could constrain the electron temperature due to the relatively soft response of \nicer, thus it was fixed to $1000$ keV. Figure \ref{fig:specres} shows two characteristic spectra representing the two spectral states that the source was in. The time evolution of the primary components is shown in Figure \ref{fig:specevo}. The top panel shows the evolution of the unabsorbed flux and the subsequent two panels display the evolution of photon index ($\Gamma$) and inner-disk temperature ($T_{in}$). The first three observations were completely disk dominated and did not require the Comptonization component. The total unabsorbed flux during these three days and the following seven days, while the source was in the HSS, is $\sim 1.2\times 10^{-7}$ erg s$^{-1}$ cm$^{-2}$ in 0.5 - 10 keV range. From 2019 January 24 onwards, when the source transitioned to the HIMS, the flux started decaying gradually till the end of observations. The contribution of the disk and Comptonization components to the total flux is marked with orange and green points respectively. It is interesting to note that, although the state transition from HSS to HIMS took place on January 24, the Comptonization component started dominating the total flux five days before that from January 19. When the observations resumed on 2019 March 18 the contribution of the individual components to the total flux had flipped, with the disk dominating the total flux, which is consistent with the state transitioning back to HSS. While in the initial HSS, the photon index varied between $4-7$ and during the HIMS it remained stable around 3. During the third phase of observations, when the source was back in HSS, $\Gamma$ had increased slightly but mostly remained unconstrained. One possible reason for this might be that the background had started dominating the high energy tail of the spectrum and starting from January 20, the spectra were fitted in the $2.5-8$ keV. Thus, for the final 20 soft state observations, $\Gamma$ was fixed to the last constrained best-fit value of 4.39 (OBSID: 2200500110). The best-fit inner-disk temperature also displays a declining trend during the HSS. It starts from $\sim 1.1$ keV on the first observation and decreases to $\sim 0.8$ keV on January 24. During the entire HIMS, $T_{in}$ varied between $0.5-0.7$ keV. Then, again during the second HSS it increased to $\sim 0.8$ keV and decreased gradually to $0.7$ keV till the end of the observations.

\section{Discussion}

We present a detailed spectral and temporal analysis of J1631 during its discovery outburst in 2018-19 with quasi-daily monitoring data from \nicer. \nicer observed J1631 for a little more than three months from 2019 January 15 to 2019 April 21. By the time \nicer commenced observation the source was already in HSS. Hence, the initial LHS and the subsequent intermediate states were missed by \nicer. This initial transition from the LHS to the HSS was so fast that it was also missed by \maxi. The left panel of Figure \ref{fig:hid} displays the full outburst HID using \maxi light curve and the right panel displays the \nicer HID \citep[also see][]{fiocchi20,monageng21}. J1631 is discovered in the LHS with the hardness lying between $1-2$ (\maxi HID) and reaches the peak at a lower hardness ratio of $\sim 0.5$. During this rising phase the source would have transitioned to the HSS. However, the transition is missed by \nicer and by the time it commences observations the source is already in the HSS. On January 24, the source transitioned to HIMS and remained there for about two months when observations were interrupted resulting in a data gap. When observations resumed on March 18, the source had again transitioned back to the HSS. Till the end of \nicer observations, the source remained in the HSS. The RID, in Figure \ref{fig:rid}, tracks the state transitions quite clearly. During the HSS the broad band variability remained around 1 per cent. The transition to the HIMS was marked by an increase in the variability to a range of $7 - 10$ per cent. As the source moved out of HIMS, the variability further decreased to the 1 per cent level.

Figure \ref{fig:specevo} shows the evolution of the primary spectral parameters during the outburst. During the initial HSS, the photon index ($\Gamma$) laid between $3 - 7$ and the inner disk temperature ($T_{in}$) ranged from $0.7 - 1.1$ keV. The HIMS was then characterised by a stable $\Gamma$ at $\sim 3$ and a $T_{in}$ that remained within $0.5 - 0.7$ keV. After transitioning back to HSS, at a lower luminosity, $T_{in}$ started from 0.8 keV and gradually decreased to 0.7 keV. The power-law index, on the other hand, could not be well constrained and was fixed to 4.39. The individual flux contribution to the total flux showed a disk dominance during the HSS and a power law dominance during the HIMS. One peculiarity being that the power-law flux started dominating 5 days before the timing properties indicate a state transition. With a limited energy range, primarily covering the soft X-rays, \nicer is not well suited to constrain the power-law component. This effect becomes more serious for J1631 as the thermal disk covers a major portion of the spectrum. This could be a possible reason behind the observed anomaly in the individual flux contributions. 

The study of temporal properties serve as a better tool for understanding state transitions. The track of the source in the RID, for example, clearly distinguishes the states. The shape and properties of the PDS are also well defined according to the spectral states. During the HSS, they have a power-law shape with very little variability (Figure \ref{fig:pds}). Being dominated by the unmodulated thermal disk component, this is expected from soft state PDSs. During the HIMS, the PDS have a flat top red noise with a moderate variability $\sim 10$ per cent. There PDS are also usually accompanied by Type-C QPOs superposed on the broadband noise.

\begin{figure}
\centering
\includegraphics[width=.95\columnwidth]{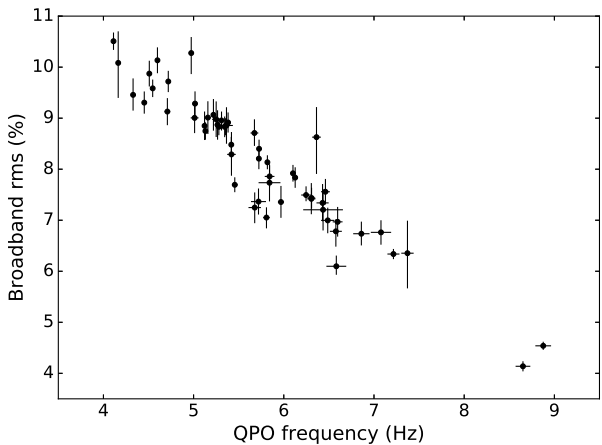}
\caption{Broadband (0.1 - 50 Hz) rms as a function of QPO frequency.}
\label{fig:nu1rms}
\end{figure}

\subsection{QPO identification and origin}

LFQPOs have been classified into three distinct types based on various properties of the PDS \citep{casella04,casella05}. Although the QPO frequency and quality factor of the three types have considerable overlap, a somewhat clear distinction arises upon comparing the shape and variability of the underlying noise component. Figure \ref{fig:nu1rms} displays the relation between the frequency of the primary QPO and the broadband rms. The anti-correlation clearly observed in this plot indicates that the QPOs are all Type-C in nature. 

\begin{figure*}
\centering
\includegraphics[scale=.5]{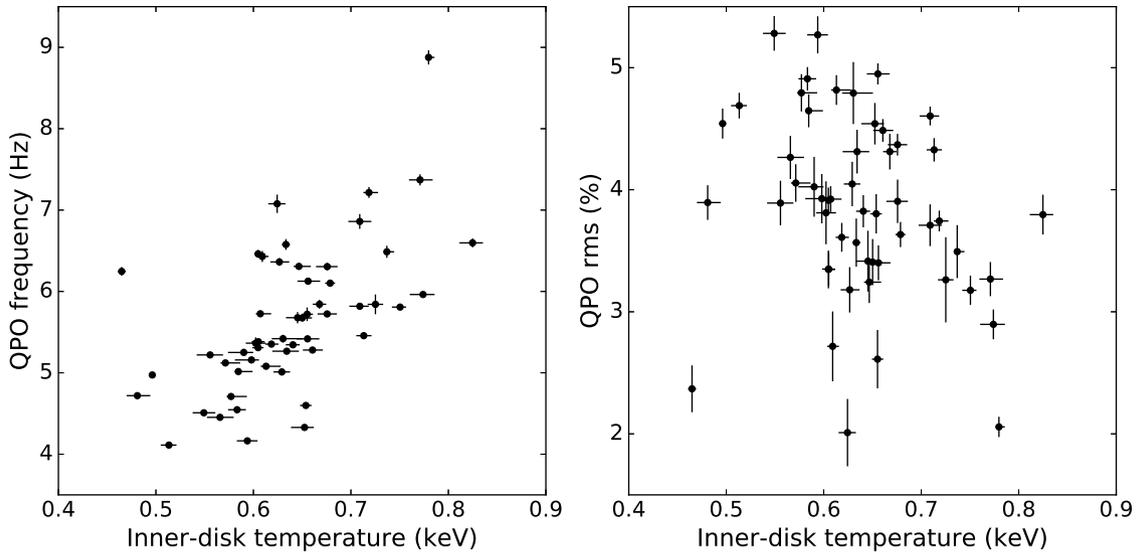}
\caption{Variation of QPO frequency (left) and rms (right) as a function of inner-disk temperature.}
\label{fig:qpotin}
\end{figure*}

Several models have been proposed to explain the origin of the LFQPOs. These are broadly classified into two categories based on the nature of origin of the QPOs, i.e., geometrical and intrinsic variability. The geometrical models of QPOs based on the Lense-Thirring precession of the inner hot flow have advanced considerably. \citet{motta15} and \citet{vandeneijnden17} have demonstrated the inclination dependence of Type-C QPOs. Type-B and A QPOs, however, are not dependent on inclination and thus could have a different origin \citep{garciaf21}. \citet{ingram09} and \citet{ingram11} have shown that the QPO frequency is strongly anti-correlated with the outer radius of the hot flow, which is nothing but the truncation radius between the thin and thick disk. This means that the inner-disk temperature, which is also anti-correlated to the inner-disk radius assuming that the accretion rate does not vary much during the HIMS (see top panel of Figure \ref{fig:specevo}), should be positively correlated with QPO frequency. The correlation of QPO frequency and rms with inner disk temperature is shown in Figure \ref{fig:qpotin}. The frequency shows a positive correlation with temperature supporting the predictions of the Lense-Thirring precession model. Since the intrinsic disk emission has little variability and QPOs originate from the hot flow, the increase in disk contribution should lead to a decrease in the rms \citep{sobolewska06, axelsson14}. Figure \ref{fig:qpotin} shows such an anti-correlation between QPO rms and $T_{in}$, although it is weak compared to the previous case. The Spearman's rank correlation coefficient for the two parameters is $\sim -0.35$ with a p-value of $\approx 0.0099$ indicating less than 1 per cent chance of being created by random noise. \citet{kalamkar15} have shown that this type of correlations are typical of BHBs in HIMS and not seen in other states. It should be noted that the decrease in count rate during the HIMS could possibly result in a corresponding decrease in mass accretion rate. However, this relation is not straightforward as the count rate also depends on the inner radius. The width of the Fe line, fitted by a \texttt{gaussian}, can be used as a proxy for the inner radius. Disks closer the black hole will experience stronger gravitational potential resulting in a broader Fe line while disks truncated at a larger distance will result in narrower Fe lines \citep{fabian00}, but also see \citet{miller06} and \citet{reis10}. We verified that $T_{in}$ is inversely correlated to the best-fit width of the Fe line and hence also the inner radius of the accretion disk.  

\subsection{rms spectra}

The study of fractional rms spectra is an important tool to help distinguishing the contribution of the individual processes and the origin of variability. \citet{gierlinski05} had studied the shape of the rms spectra for various spectral states and deciphered the relative contributions of the various components to the total variability \citep[see also][for a similar description with a slightly different state classification scheme]{belloni11}. While flat rms spectra indicate variability in the normalization of the Comptonization component, hard and inverted spectra would require the variation of spectral shape along with normalisation \citep{gierlinski05}. The rms spectra of J1631 is displayed in Figure \ref{fig:rmsspec}. During the HSS, almost all the energy bands have very little variability and the amplitude could not be constrained. This is not surprising considering that even the energy integrated PDS (see Figure \ref{fig:pds}) during the HSS had almost no power. The rms spectra during the HIMS have a hard shape that is characteristic of a late state HIMS which is about to transition to the HSS \citep{belloni11}. As mentioned above, this shape would arise because of variability in the Compton tail along with normalization of the Compton spectrum. The first two energy bands (E1 and E2) during the HIMS could not be constrained for most of the observations. Some of the best case scenarios are shown in Figure \ref{fig:rmsspec}, particularly in Panels (d), (e), (f), and (g). Since we only have upper limits for the first two energy bands, the true shape is uncertain. It could either be flat as expected for the very high state (VHS) cases studied by \citet{gierlinski05}, or could have an inverted shape with a soft excess. One clear example of this kind of spectrum is shown in Panel (d). Panels (e) and (g) could also fall in this category, but we can not be certain. While the minima in the E3 (or sometimes E2) energy band can be understood as a dominant contribution from the unmodulated thermal disk, the excess in the E1 and E2 bands is puzzling.

\section{Summary}

In this work we present a comprehensive spectral and timing study of the black-hole transient J1631 during its 2018-19 outburst using observations from the \nicer observatory. The main results are noted below. 

\begin{itemize}
    \item \nicer observed J1631 for about three months tracking the decay phase of the outburst. The observations began while the source was in the HSS and after that the source moved to the HIMS for a couple of months. The transition back to the HSS was missed by \nicer and the source continued to decay with decreasing hardness till the end of the outburst. Both the beginning and end LHS of the source were missed by \nicer.
    \item The spectra could be fitted by a combination of a multicolor blackbody and a thermal Comptonization component. During the bright phases, a Gaussian was required to account for the fluorescent Fe line. 
    \item During the HSS, the variability was very low ($\sim 1$ per cent) and the PDS was a featureless power law. The PDS in the HIMS had a broadband noise component along with peaked noises and QPOs. All QPOs were of Type-C in nature. 
    \item The frequency of the QPOs increase and QPO rms decrease with increasing inner-disk temperature. 
    \item The rms spectrum of J1631 in the HIMS is hard above 1 keV. The shape of the spectra below 1 keV is uncertain because the power in the first two bands could not be constrained. Some of the observations have more power in the lower energy bands. This effect is puzzling and not explained by merely varying the normalization and power-law slope of the Compton spectrum.
\end{itemize}

\section*{Acknowledgements}

SKR and SV acknowledge the support of Physical Research Laboratory which is funded by Department of Space, Government of India. TMB acknowledges support from the agreement ASI-INAF n.2017-14-H.0 and PRIN-INAF 2019 n.15. This research has made use of data and software provided by the High Energy Astrophysics Science Archive Research Center (HEASARC), which is a service of the Astrophysics Science Division at NASA/GSFC. It has also made use of the MAXI data provided by RIKEN, JAXA and the MAXI team. 

\section*{Data Availability}

The data underlying this article are available in the HEASARC archives.   



\bibliographystyle{mnras}
\bibliography{j1631_nicer} 

\begin{thebibliography}{}
\makeatletter
\relax
\def\mn@urlcharsother{\let\do\@makeother \do\$\do\&\do\#\do\^\do\_\do\%\do\~}
\def\mn@doi{\begingroup\mn@urlcharsother \@ifnextchar [ {\mn@doi@}
  {\mn@doi@[]}}
\def\mn@doi@[#1]#2{\def\@tempa{#1}\ifx\@tempa\@empty \href
  {http://dx.doi.org/#2} {doi:#2}\else \href {http://dx.doi.org/#2} {#1}\fi
  \endgroup}
\def\mn@eprint#1#2{\mn@eprint@#1:#2::\@nil}
\def\mn@eprint@arXiv#1{\href {http://arxiv.org/abs/#1} {{\tt arXiv:#1}}}
\def\mn@eprint@dblp#1{\href {http://dblp.uni-trier.de/rec/bibtex/#1.xml}
  {dblp:#1}}
\def\mn@eprint@#1:#2:#3:#4\@nil{\def\@tempa {#1}\def\@tempb {#2}\def\@tempc
  {#3}\ifx \@tempc \@empty \let \@tempc \@tempb \let \@tempb \@tempa \fi \ifx
  \@tempb \@empty \def\@tempb {arXiv}\fi \@ifundefined
  {mn@eprint@\@tempb}{\@tempb:\@tempc}{\expandafter \expandafter \csname
  mn@eprint@\@tempb\endcsname \expandafter{\@tempc}}}

\bibitem[\protect\citeauthoryear{{Arnaud}}{{Arnaud}}{1996}]{arnaud96}
{Arnaud} K.~A.,  1996, in {Jacoby} G.~H.,  {Barnes} J.,  eds,  Astronomical
  Society of the Pacific Conference Series Vol. 101, Astronomical Data Analysis
  Software and Systems V. p.~17

\bibitem[\protect\citeauthoryear{{Axelsson}, {Done}  \&
  {Hjalmarsdotter}}{{Axelsson} et~al.}{2014}]{axelsson14}
{Axelsson} M.,  {Done} C.,   {Hjalmarsdotter} L.,  2014, \mn@doi [\mnras]
  {10.1093/mnras/stt2236}, \href
  {https://ui.adsabs.harvard.edu/abs/2014MNRAS.438..657A} {438, 657}

\bibitem[\protect\citeauthoryear{{Belloni}}{{Belloni}}{2010}]{belloni10}
{Belloni} T.~M.,  2010, {States and Transitions in Black Hole Binaries}.
p.~53, \mn@doi{10.1007/978-3-540-76937-8_3}

\bibitem[\protect\citeauthoryear{{Belloni} \& {Hasinger}}{{Belloni} \&
  {Hasinger}}{1990}]{belloni90}
{Belloni} T.,  {Hasinger} G.,  1990, \aap, \href
  {https://ui.adsabs.harvard.edu/abs/1990A&A...230..103B} {230, 103}

\bibitem[\protect\citeauthoryear{{Belloni} \& {Motta}}{{Belloni} \&
  {Motta}}{2016}]{belloni16}
{Belloni} T.~M.,  {Motta} S.~E.,  2016, {Transient Black Hole Binaries}.
Springer, p.~61, \mn@doi{10.1007/978-3-662-52859-4_2}

\bibitem[\protect\citeauthoryear{{Belloni}, {Psaltis}  \& {van der
  Klis}}{{Belloni} et~al.}{2002}]{belloni02}
{Belloni} T.,  {Psaltis} D.,   {van der Klis} M.,  2002, \mn@doi [\apj]
  {10.1086/340290}, \href
  {https://ui.adsabs.harvard.edu/abs/2002ApJ...572..392B} {572, 392}

\bibitem[\protect\citeauthoryear{{Belloni}, {Motta}  \&
  {Mu{\~n}oz-Darias}}{{Belloni} et~al.}{2011}]{belloni11}
{Belloni} T.~M.,  {Motta} S.~E.,   {Mu{\~n}oz-Darias} T.,  2011, Bulletin of
  the Astronomical Society of India, \href
  {https://ui.adsabs.harvard.edu/abs/2011BASI...39..409B} {39, 409}

\bibitem[\protect\citeauthoryear{{Cabanac}, {Henri}, {Petrucci}, {Malzac},
  {Ferreira}  \& {Belloni}}{{Cabanac} et~al.}{2010}]{cabanac10}
{Cabanac} C.,  {Henri} G.,  {Petrucci} P.~O.,  {Malzac} J.,  {Ferreira} J.,
  {Belloni} T.~M.,  2010, \mn@doi [\mnras] {10.1111/j.1365-2966.2010.16340.x},
  \href {https://ui.adsabs.harvard.edu/abs/2010MNRAS.404..738C} {404, 738}

\bibitem[\protect\citeauthoryear{{Casella}, {Belloni}, {Homan}  \&
  {Stella}}{{Casella} et~al.}{2004}]{casella04}
{Casella} P.,  {Belloni} T.,  {Homan} J.,   {Stella} L.,  2004, \mn@doi [\aap]
  {10.1051/0004-6361:20041231}, \href
  {https://ui.adsabs.harvard.edu/abs/2004A&A...426..587C} {426, 587}

\bibitem[\protect\citeauthoryear{{Casella}, {Belloni}  \& {Stella}}{{Casella}
  et~al.}{2005}]{casella05}
{Casella} P.,  {Belloni} T.,   {Stella} L.,  2005, \mn@doi [\apj]
  {10.1086/431174}, \href
  {https://ui.adsabs.harvard.edu/abs/2005ApJ...629..403C} {629, 403}

\bibitem[\protect\citeauthoryear{{Chakrabarti}, {Debnath}, {Nandi}  \&
  {Pal}}{{Chakrabarti} et~al.}{2008}]{chakrabarti08}
{Chakrabarti} S.~K.,  {Debnath} D.,  {Nandi} A.,   {Pal} P.~S.,  2008, \mn@doi
  [\aap] {10.1051/0004-6361:200810136}, \href
  {https://ui.adsabs.harvard.edu/abs/2008A&A...489L..41C} {489, L41}

\bibitem[\protect\citeauthoryear{{Done}, {Gierli{\'n}ski}  \& {Kubota}}{{Done}
  et~al.}{2007}]{done07}
{Done} C.,  {Gierli{\'n}ski} M.,   {Kubota} A.,  2007, \mn@doi [\aapr]
  {10.1007/s00159-007-0006-1}, \href
  {https://ui.adsabs.harvard.edu/abs/2007A&ARv..15....1D} {15, 1}

\bibitem[\protect\citeauthoryear{{Fabian}, {Iwasawa}, {Reynolds}  \&
  {Young}}{{Fabian} et~al.}{2000}]{fabian00}
{Fabian} A.~C.,  {Iwasawa} K.,  {Reynolds} C.~S.,   {Young} A.~J.,  2000,
  \mn@doi [\pasp] {10.1086/316610}, \href
  {https://ui.adsabs.harvard.edu/abs/2000PASP..112.1145F} {112, 1145}

\bibitem[\protect\citeauthoryear{{Fiocchi} et~al.,}{{Fiocchi}
  et~al.}{2020}]{fiocchi20}
{Fiocchi} M.,  et~al., 2020, \mn@doi [\mnras] {10.1093/mnras/staa068}, \href
  {https://ui.adsabs.harvard.edu/abs/2020MNRAS.492.3657F} {492, 3657}

\bibitem[\protect\citeauthoryear{{Garc{\'\i}a}, {M{\'e}ndez}, {Karpouzas},
  {Belloni}, {Zhang}  \& {Altamirano}}{{Garc{\'\i}a} et~al.}{2021}]{garciaf21}
{Garc{\'\i}a} F.,  {M{\'e}ndez} M.,  {Karpouzas} K.,  {Belloni} T.,  {Zhang}
  L.,   {Altamirano} D.,  2021, \mn@doi [\mnras] {10.1093/mnras/staa3944},
  \href {https://ui.adsabs.harvard.edu/abs/2021MNRAS.501.3173G} {501, 3173}

\bibitem[\protect\citeauthoryear{{Gendreau}, {Arzoumanian}  \&
  {Okajima}}{{Gendreau} et~al.}{2012}]{gendreau12}
{Gendreau} K.~C.,  {Arzoumanian} Z.,   {Okajima} T.,  2012, in Space Telescopes
  and Instrumentation 2012: Ultraviolet to Gamma Ray. p. 844313,
  \mn@doi{10.1117/12.926396}

\bibitem[\protect\citeauthoryear{{Gierli{\'n}ski} \&
  {Zdziarski}}{{Gierli{\'n}ski} \& {Zdziarski}}{2005}]{gierlinski05}
{Gierli{\'n}ski} M.,  {Zdziarski} A.~A.,  2005, \mn@doi [\mnras]
  {10.1111/j.1365-2966.2005.09527.x}, \href
  {https://ui.adsabs.harvard.edu/abs/2005MNRAS.363.1349G} {363, 1349}

\bibitem[\protect\citeauthoryear{{Gilfanov}}{{Gilfanov}}{2010}]{gilfanov10}
{Gilfanov} M.,  2010, {X-Ray Emission from Black-Hole Binaries}.
p.~17, \mn@doi{10.1007/978-3-540-76937-8_2}

\bibitem[\protect\citeauthoryear{{Heil}, {Uttley}  \& {Klein-Wolt}}{{Heil}
  et~al.}{2015}]{heil15}
{Heil} L.~M.,  {Uttley} P.,   {Klein-Wolt} M.,  2015, \mn@doi [\mnras]
  {10.1093/mnras/stv240}, \href
  {https://ui.adsabs.harvard.edu/abs/2015MNRAS.448.3348H} {448, 3348}

\bibitem[\protect\citeauthoryear{{Homan} \& {Belloni}}{{Homan} \&
  {Belloni}}{2005}]{homan05}
{Homan} J.,  {Belloni} T.,  2005, \mn@doi [\apss] {10.1007/s10509-005-1197-4},
  \href {https://ui.adsabs.harvard.edu/abs/2005Ap&SS.300..107H} {300, 107}

\bibitem[\protect\citeauthoryear{{Homan}, {Wijnands}, {van der Klis},
  {Belloni}, {van Paradijs}, {Klein-Wolt}, {Fender}  \& {M{\'e}ndez}}{{Homan}
  et~al.}{2001}]{homan01}
{Homan} J.,  {Wijnands} R.,  {van der Klis} M.,  {Belloni} T.,  {van Paradijs}
  J.,  {Klein-Wolt} M.,  {Fender} R.,   {M{\'e}ndez} M.,  2001, \mn@doi [\apjs]
  {10.1086/318954}, \href
  {https://ui.adsabs.harvard.edu/abs/2001ApJS..132..377H} {132, 377}

\bibitem[\protect\citeauthoryear{{Ingram} \& {Done}}{{Ingram} \&
  {Done}}{2011}]{ingram11}
{Ingram} A.,  {Done} C.,  2011, \mn@doi [\mnras]
  {10.1111/j.1365-2966.2011.18860.x}, \href
  {https://ui.adsabs.harvard.edu/abs/2011MNRAS.415.2323I} {415, 2323}

\bibitem[\protect\citeauthoryear{{Ingram}, {Done}  \& {Fragile}}{{Ingram}
  et~al.}{2009}]{ingram09}
{Ingram} A.,  {Done} C.,   {Fragile} P.~C.,  2009, \mn@doi [\mnras]
  {10.1111/j.1745-3933.2009.00693.x}, \href
  {https://ui.adsabs.harvard.edu/abs/2009MNRAS.397L.101I} {397, L101}

\bibitem[\protect\citeauthoryear{{Ingram}, {van der Klis}, {Middleton}, {Done},
  {Altamirano}, {Heil}, {Uttley}  \& {Axelsson}}{{Ingram}
  et~al.}{2016}]{ingram16}
{Ingram} A.,  {van der Klis} M.,  {Middleton} M.,  {Done} C.,  {Altamirano} D.,
   {Heil} L.,  {Uttley} P.,   {Axelsson} M.,  2016, \mn@doi [\mnras]
  {10.1093/mnras/stw1245}, \href
  {https://ui.adsabs.harvard.edu/abs/2016MNRAS.461.1967I} {461, 1967}

\bibitem[\protect\citeauthoryear{{Ingram}, {van der Klis}, {Middleton},
  {Altamirano}  \& {Uttley}}{{Ingram} et~al.}{2017}]{ingram17}
{Ingram} A.,  {van der Klis} M.,  {Middleton} M.,  {Altamirano} D.,   {Uttley}
  P.,  2017, \mn@doi [\mnras] {10.1093/mnras/stw2581}, \href
  {https://ui.adsabs.harvard.edu/abs/2017MNRAS.464.2979I} {464, 2979}

\bibitem[\protect\citeauthoryear{{Kalamkar}, {Reynolds}, {van der Klis},
  {Altamirano}  \& {Miller}}{{Kalamkar} et~al.}{2015}]{kalamkar15}
{Kalamkar} M.,  {Reynolds} M.~T.,  {van der Klis} M.,  {Altamirano} D.,
  {Miller} J.~M.,  2015, \mn@doi [\apj] {10.1088/0004-637X/802/1/23}, \href
  {https://ui.adsabs.harvard.edu/abs/2015ApJ...802...23K} {802, 23}

\bibitem[\protect\citeauthoryear{{Karpouzas}, {M{\'e}ndez}, {Ribeiro},
  {Altamirano}, {Blaes}  \& {Garc{\'\i}a}}{{Karpouzas}
  et~al.}{2020}]{karpouzas20}
{Karpouzas} K.,  {M{\'e}ndez} M.,  {Ribeiro} E.~M.,  {Altamirano} D.,  {Blaes}
  O.,   {Garc{\'\i}a} F.,  2020, \mn@doi [\mnras] {10.1093/mnras/stz3502},
  \href {https://ui.adsabs.harvard.edu/abs/2020MNRAS.492.1399K} {492, 1399}

\bibitem[\protect\citeauthoryear{{Kumar} \& {Misra}}{{Kumar} \&
  {Misra}}{2014}]{kumar14}
{Kumar} N.,  {Misra} R.,  2014, \mn@doi [\mnras] {10.1093/mnras/stu1946}, \href
  {https://ui.adsabs.harvard.edu/abs/2014MNRAS.445.2818K} {445, 2818}

\bibitem[\protect\citeauthoryear{{Lasota}}{{Lasota}}{2001}]{lasota01}
{Lasota} J.-P.,  2001, \mn@doi [\nar] {10.1016/S1387-6473(01)00112-9}, \href
  {https://ui.adsabs.harvard.edu/abs/2001NewAR..45..449L} {45, 449}

\bibitem[\protect\citeauthoryear{{Lee} \& {Miller}}{{Lee} \&
  {Miller}}{1998}]{lee98}
{Lee} H.~C.,  {Miller} G.~S.,  1998, \mn@doi [\mnras]
  {10.1046/j.1365-8711.1998.01842.x}, \href
  {https://ui.adsabs.harvard.edu/abs/1998MNRAS.299..479L} {299, 479}

\bibitem[\protect\citeauthoryear{{Matsuoka} et~al.,}{{Matsuoka}
  et~al.}{2009}]{matsuoka09}
{Matsuoka} M.,  et~al., 2009, \mn@doi [PASJ] {10.1093/pasj/61.5.999}, \href
  {https://ui.adsabs.harvard.edu/abs/2009PASJ...61..999M} {61, 999}

\bibitem[\protect\citeauthoryear{{Miller}, {Homan}, {Steeghs}, {Rupen},
  {Hunstead}, {Wijnands}, {Charles}  \& {Fabian}}{{Miller}
  et~al.}{2006}]{miller06}
{Miller} J.~M.,  {Homan} J.,  {Steeghs} D.,  {Rupen} M.,  {Hunstead} R.~W.,
  {Wijnands} R.,  {Charles} P.~A.,   {Fabian} A.~C.,  2006, \mn@doi [\apj]
  {10.1086/508644}, \href
  {https://ui.adsabs.harvard.edu/abs/2006ApJ...653..525M} {653, 525}

\bibitem[\protect\citeauthoryear{{Miyamoto}, {Iga}, {Kitamoto}  \&
  {Kamado}}{{Miyamoto} et~al.}{1993}]{miyamoto93}
{Miyamoto} S.,  {Iga} S.,  {Kitamoto} S.,   {Kamado} Y.,  1993, \mn@doi [\apjl]
  {10.1086/186716}, \href
  {https://ui.adsabs.harvard.edu/abs/1993ApJ...403L..39M} {403, L39}

\bibitem[\protect\citeauthoryear{{Miyasaka}, {Tomsick}, {Xu}  \&
  {Harrison}}{{Miyasaka} et~al.}{2018}]{miyasaka18}
{Miyasaka} H.,  {Tomsick} J.~A.,  {Xu} Y.,   {Harrison} F.~A.,  2018, The
  Astronomer's Telegram, \href
  {https://ui.adsabs.harvard.edu/abs/2018ATel12340....1M} {12340, 1}

\bibitem[\protect\citeauthoryear{{Molteni}, {Sponholz}  \&
  {Chakrabarti}}{{Molteni} et~al.}{1996}]{molteni96}
{Molteni} D.,  {Sponholz} H.,   {Chakrabarti} S.~K.,  1996, \mn@doi [\apj]
  {10.1086/176775}, \href
  {https://ui.adsabs.harvard.edu/abs/1996ApJ...457..805M} {457, 805}

\bibitem[\protect\citeauthoryear{{Monageng}, {Motta}, {Fender}, {Yu}, {Woudt},
  {Tremou}, {Miller-Jones}  \& {van der Horst}}{{Monageng}
  et~al.}{2021}]{monageng21}
{Monageng} I.~M.,  {Motta} S.~E.,  {Fender} R.,  {Yu} W.,  {Woudt} P.~A.,
  {Tremou} E.,  {Miller-Jones} J.~C.~A.,   {van der Horst} A.~J.,  2021,
  \mn@doi [\mnras] {10.1093/mnras/stab043}, \href
  {https://ui.adsabs.harvard.edu/abs/2021MNRAS.501.5776M} {501, 5776}

\bibitem[\protect\citeauthoryear{{Motta}, {Casella}, {Henze},
  {Mu{\~n}oz-Darias}, {Sanna}, {Fender}  \& {Belloni}}{{Motta}
  et~al.}{2015}]{motta15}
{Motta} S.~E.,  {Casella} P.,  {Henze} M.,  {Mu{\~n}oz-Darias} T.,  {Sanna} A.,
   {Fender} R.,   {Belloni} T.,  2015, \mn@doi [\mnras]
  {10.1093/mnras/stu2579}, \href
  {https://ui.adsabs.harvard.edu/abs/2015MNRAS.447.2059M} {447, 2059}

\bibitem[\protect\citeauthoryear{{Mu{\~n}oz-Darias}, {Motta}  \&
  {Belloni}}{{Mu{\~n}oz-Darias} et~al.}{2011}]{munozdarias11}
{Mu{\~n}oz-Darias} T.,  {Motta} S.,   {Belloni} T.~M.,  2011, \mn@doi [\mnras]
  {10.1111/j.1365-2966.2010.17476.x}, \href
  {https://ui.adsabs.harvard.edu/abs/2011MNRAS.410..679M} {410, 679}

\bibitem[\protect\citeauthoryear{{O'Neill}, {Reynolds}, {Miller}  \&
  {Sorathia}}{{O'Neill} et~al.}{2011}]{oneill11}
{O'Neill} S.~M.,  {Reynolds} C.~S.,  {Miller} M.~C.,   {Sorathia} K.~A.,  2011,
  \mn@doi [\apj] {10.1088/0004-637X/736/2/107}, \href
  {https://ui.adsabs.harvard.edu/abs/2011ApJ...736..107O} {736, 107}

\bibitem[\protect\citeauthoryear{{Reis}, {Fabian}  \& {Miller}}{{Reis}
  et~al.}{2010}]{reis10}
{Reis} R.~C.,  {Fabian} A.~C.,   {Miller} J.~M.,  2010, \mn@doi [\mnras]
  {10.1111/j.1365-2966.2009.15976.x}, \href
  {https://ui.adsabs.harvard.edu/abs/2010MNRAS.402..836R} {402, 836}

\bibitem[\protect\citeauthoryear{{Remillard} \& {McClintock}}{{Remillard} \&
  {McClintock}}{2006}]{remillard06}
{Remillard} R.~A.,  {McClintock} J.~E.,  2006, \mn@doi [\araa]
  {10.1146/annurev.astro.44.051905.092532}, \href
  {https://ui.adsabs.harvard.edu/abs/2006ARA&A..44...49R} {44, 49}

\bibitem[\protect\citeauthoryear{{Schnittman}, {Homan}  \&
  {Miller}}{{Schnittman} et~al.}{2006}]{schnittman06}
{Schnittman} J.~D.,  {Homan} J.,   {Miller} J.~M.,  2006, \mn@doi [\apj]
  {10.1086/500923}, \href
  {https://ui.adsabs.harvard.edu/abs/2006ApJ...642..420S} {642, 420}

\bibitem[\protect\citeauthoryear{{Sobolewska} \& {{\.Z}ycki}}{{Sobolewska} \&
  {{\.Z}ycki}}{2006}]{sobolewska06}
{Sobolewska} M.~A.,  {{\.Z}ycki} P.~T.,  2006, \mn@doi [\mnras]
  {10.1111/j.1365-2966.2006.10489.x}, \href
  {https://ui.adsabs.harvard.edu/abs/2006MNRAS.370..405S} {370, 405}

\bibitem[\protect\citeauthoryear{{Stella} \& {Vietri}}{{Stella} \&
  {Vietri}}{1998}]{stella98}
{Stella} L.,  {Vietri} M.,  1998, \mn@doi [\apjl] {10.1086/311075}, \href
  {https://ui.adsabs.harvard.edu/abs/1998ApJ...492L..59S} {492, L59}

\bibitem[\protect\citeauthoryear{{Stella}, {Vietri}  \& {Morsink}}{{Stella}
  et~al.}{1999}]{stella99}
{Stella} L.,  {Vietri} M.,   {Morsink} S.~M.,  1999, \mn@doi [\apjl]
  {10.1086/312291}, \href
  {https://ui.adsabs.harvard.edu/abs/1999ApJ...524L..63S} {524, L63}

\bibitem[\protect\citeauthoryear{{Tagger} \& {Pellat}}{{Tagger} \&
  {Pellat}}{1999}]{tagger99}
{Tagger} M.,  {Pellat} R.,  1999, \aap, \href
  {https://ui.adsabs.harvard.edu/abs/1999A&A...349.1003T} {349, 1003}

\bibitem[\protect\citeauthoryear{{Titarchuk} \& {Fiorito}}{{Titarchuk} \&
  {Fiorito}}{2004}]{titarchuk04}
{Titarchuk} L.,  {Fiorito} R.,  2004, \mn@doi [\apj] {10.1086/422573}, \href
  {https://ui.adsabs.harvard.edu/abs/2004ApJ...612..988T} {612, 988}

\bibitem[\protect\citeauthoryear{{Verner}, {Ferland}, {Korista}  \&
  {Yakovlev}}{{Verner} et~al.}{1996}]{verner96}
{Verner} D.~A.,  {Ferland} G.~J.,  {Korista} K.~T.,   {Yakovlev} D.~G.,  1996,
  \mn@doi [\apj] {10.1086/177435}, \href
  {https://ui.adsabs.harvard.edu/abs/1996ApJ...465..487V} {465, 487}

\bibitem[\protect\citeauthoryear{{Wagoner}, {Silbergleit}  \&
  {Ortega-Rodr{\'\i}guez}}{{Wagoner} et~al.}{2001}]{wagoner01}
{Wagoner} R.~V.,  {Silbergleit} A.~S.,   {Ortega-Rodr{\'\i}guez} M.,  2001,
  \mn@doi [\apjl] {10.1086/323655}, \href
  {https://ui.adsabs.harvard.edu/abs/2001ApJ...559L..25W} {559, L25}

\bibitem[\protect\citeauthoryear{{Wilms}, {Allen}  \& {McCray}}{{Wilms}
  et~al.}{2000}]{wilms00}
{Wilms} J.,  {Allen} A.,   {McCray} R.,  2000, \mn@doi [\apj] {10.1086/317016},
  \href {https://ui.adsabs.harvard.edu/abs/2000ApJ...542..914W} {542, 914}

\bibitem[\protect\citeauthoryear{{Xu}, {Harrison}, {Tomsick}, {Walton},
  {Barret}, {Garc{\'\i}a}, {Hare}  \& {Parker}}{{Xu} et~al.}{2020}]{xu20}
{Xu} Y.,  {Harrison} F.~A.,  {Tomsick} J.~A.,  {Walton} D.~J.,  {Barret} D.,
  {Garc{\'\i}a} J.~A.,  {Hare} J.,   {Parker} M.~L.,  2020, \mn@doi [\apj]
  {10.3847/1538-4357/ab7dc0}, \href
  {https://ui.adsabs.harvard.edu/abs/2020ApJ...893...30X} {893, 30}

\bibitem[\protect\citeauthoryear{{Zdziarski}, {Johnson}  \&
  {Magdziarz}}{{Zdziarski} et~al.}{1996}]{zdziarski96}
{Zdziarski} A.~A.,  {Johnson} W.~N.,   {Magdziarz} P.,  1996, \mn@doi [MNRAS]
  {10.1093/mnras/283.1.193}, \href
  {https://ui.adsabs.harvard.edu/abs/1996MNRAS.283..193Z} {283, 193}

\bibitem[\protect\citeauthoryear{{{\.Z}ycki}, {Done}  \& {Smith}}{{{\.Z}ycki}
  et~al.}{1999}]{zycki99}
{{\.Z}ycki} P.~T.,  {Done} C.,   {Smith} D.~A.,  1999, \mn@doi [MNRAS]
  {10.1046/j.1365-8711.1999.02885.x}, \href
  {https://ui.adsabs.harvard.edu/abs/1999MNRAS.309..561Z} {309, 561}

\bibitem[\protect\citeauthoryear{{van den Eijnden}, {Ingram}, {Uttley},
  {Motta}, {Belloni}  \& {Gardenier}}{{van den Eijnden}
  et~al.}{2017}]{vandeneijnden17}
{van den Eijnden} J.,  {Ingram} A.,  {Uttley} P.,  {Motta} S.~E.,  {Belloni}
  T.~M.,   {Gardenier} D.~W.,  2017, \mn@doi [\mnras] {10.1093/mnras/stw2634},
  \href {https://ui.adsabs.harvard.edu/abs/2017MNRAS.464.2643V} {464, 2643}

\makeatother
\end{thebibliography}

\bsp	
\label{lastpage}
\end{document}